\documentclass[aps,prb,twocolumn,superscriptaddress,floatfix]{revtex4-1}

\usepackage{graphicx}
\usepackage{color}
\usepackage{ifsym}
\usepackage{amssymb}
\usepackage{bm}

\usepackage[ bf, sf, justification=raggedright]{caption}
\DeclareCaptionLabelSeparator{line}{ $|$ }
\usepackage{xr}
\newcommand{\FeCoSi}{F\lowercase{e}$_{1-x}$C\lowercase{o}$_{x}$S\lowercase{i}}

\begin{document}

\title{Giant topological Hall effect in strained Fe$_{0.7}$Co$_{0.3}$Si epilayers}

\author{Nicholas~A.~Porter}\email[email:~]{n.a.porter@leeds.ac.uk}
\affiliation{School of Physics \&\ Astronomy, University of Leeds, Leeds LS2 9JT, United Kingdom}

\author{Priyasmita~Sinha}
\affiliation{School of Physics \&\ Astronomy, University of Leeds, Leeds LS2 9JT, United Kingdom}

\author{Michael~B.~Ward}
\affiliation{Institute for Materials Research, SPEME, University of Leeds, Leeds LS2 9JT, United Kingdom}

\author{Alexey~N.~Dobrynin}
\affiliation{Diamond Light Source, Chilton, Didcot, Oxon. OX11 0DE, United Kingdom}

\author{Rik~M.~D.~Brydson}
\affiliation{Institute for Materials Research, SPEME, University of Leeds, Leeds LS2 9JT, United Kingdom}

\author{Timothy~R.~Charlton}
\affiliation{ISIS, STFC Rutherford Appleton Laboratory, Chilton, Didcot, Oxon. OX11 0QX, United Kingdom}

\author{Christian~J.~Kinane}
\affiliation{ISIS, STFC Rutherford Appleton Laboratory, Chilton, Didcot, Oxon. OX11 0QX, United Kingdom}

\author{Michael~D.~Robertson}
\affiliation{Department of Physics, Acadia University, Wolfville, Nova Scotia, B4P 2R6, Canada}

\author{Sean~Langridge}
\affiliation{ISIS, STFC Rutherford Appleton Laboratory, Chilton, Didcot, Oxon. OX11 0QX, United Kingdom}

\author{Christopher~H.~Marrows}\email[email:~]{c.h.marrows@leeds.ac.uk}
\affiliation{School of Physics \&\ Astronomy, University of Leeds, Leeds LS2 9JT, United Kingdom}

\begin{abstract}
  The coupling of electron spin to real-space magnetic textures leads to a variety of interesting magnetotransport effects. The skyrmionic spin textures often found in chiral B20-lattice magnets give rise, via real-space Berry phases, to the topological Hall effect, but it is typically rather small. Here, B20-ordered Fe$_{0.7}$Co$_{0.3}$Si epilayers display a giant topological Hall effect due to the combination of three favourable properties: they have a high spin-polarisation, a large ordinary Hall coefficient, and dense chiral spin textures. The topological Hall resistivity is as large as 820~n$\Omega$cm at helium temperatures. Moreover, we observed a drop in the longitudinal resistivity of 100~n$\Omega$cm at low temperatures in the same field range, suggesting that it is also of topological origin. That such strong effects can be found in material grown in thin film form on commercial silicon wafer bodes well for skyrmion-based spintronics.
\end{abstract}


\date{\today}
\maketitle

Quantum mechanical Berry phases in magnetic materials give rise to changes in electron motion. Berry curvature of the band structure in reciprocal space leads to the intrinsic part of the anomalous Hall effect\cite{Nagaosa2010}. Meanwhile, as electrons traverse a spin texture they accumulate Berry phase in real space, leading to the topological Hall effect\cite{Bruno2004,Tatara2008,Nagaosa2012a}. Recently, examples of non-trivial spin textures have been found in magnetic crystals with the chiral B20 lattice\cite{Muhlbauer2009,Yu2010,Yu2011,Tonomura2012,Wilhelm2011,Adams2011,Munzer2010,Munzer2010}, which have a non-zero Dzyaloshinskii-Moriya interaction (DMI), leading to a helical ground state and skyrmion phases\cite{Bogdanov2001,Rossler2006}. The skyrmions are topologically stable localised defects in the magnetisation, each of which imparts a quantised Berry phase of $4\pi$ to an electron that traverses it. Hence, the appearance of a topological Hall effect is a signature of the presence of skyrmions, but it is typically rather small\cite{Lee2009,Neubauer2009,Schulz2012}. Here we show that epitaxially strained but B20-ordered Fe$_{0.7}$Co$_{0.3}$Si thin films display a very large topological Hall effect. The topological Hall resistivity is $\sim 820$~n$\Omega$cm at helium temperatures, more than two orders of magnitude larger than that found in MnSi when the effect was initially discovered in this class of materials (4~n$\Omega$cm)\cite{Neubauer2009}, and roughly five times as large as the effect seen in MnGe (160~n$\Omega$cm)\cite{Kanazawa2011}. Moreover, we observe a contribution to the longitudinal resistivity of 100~n$\Omega$cm at the lowest temperatures whilst within the topologically non-trivial phase, potentially also with its origin in Berry phase effects\cite{Yi2009}.

The \FeCoSi\ system is fascinating for many reasons. Neither endmember displays magnetic order. FeSi is a narrow-gap semiconductor with anomalous paramagnetism at high temperatures\cite{Jaccarino1967} that shares many features with Kondo insulators\cite{Aeppli1992}, whilst CoSi is a diamagnetic metal\cite{Wernick1972}. Nevertheless, magnetic order is present for almost all intermediate compositions\cite{Wernick1972,Manyala2000}. The unusual band structure leads to an unusual non-saturating linear magnetoresistance at high fields\cite{Manyala2000,Onose2005,Porter2012}, and the material is predicted to be half-metallic for compositions in the range $0 < x \lesssim 0.25$\cite{Guevara2004}. Indeed, results close to the expected one Bohr magneton per electron are found in this range\cite{Manyala2000,Morley2011,Sinha2013}. The chiral B20 crystal structure leads to a helimagnetic ground state\cite{Beille1981,Uchida2006} and a skyrmion-bearing phase for appropriate values of magnetic field and temperature\cite{Munzer2010,Yu2010}.

\section*{Growth and characterisation of \FeCoSi\ epilayers}

\begin{figure}[t]
  \includegraphics[width=8cm]{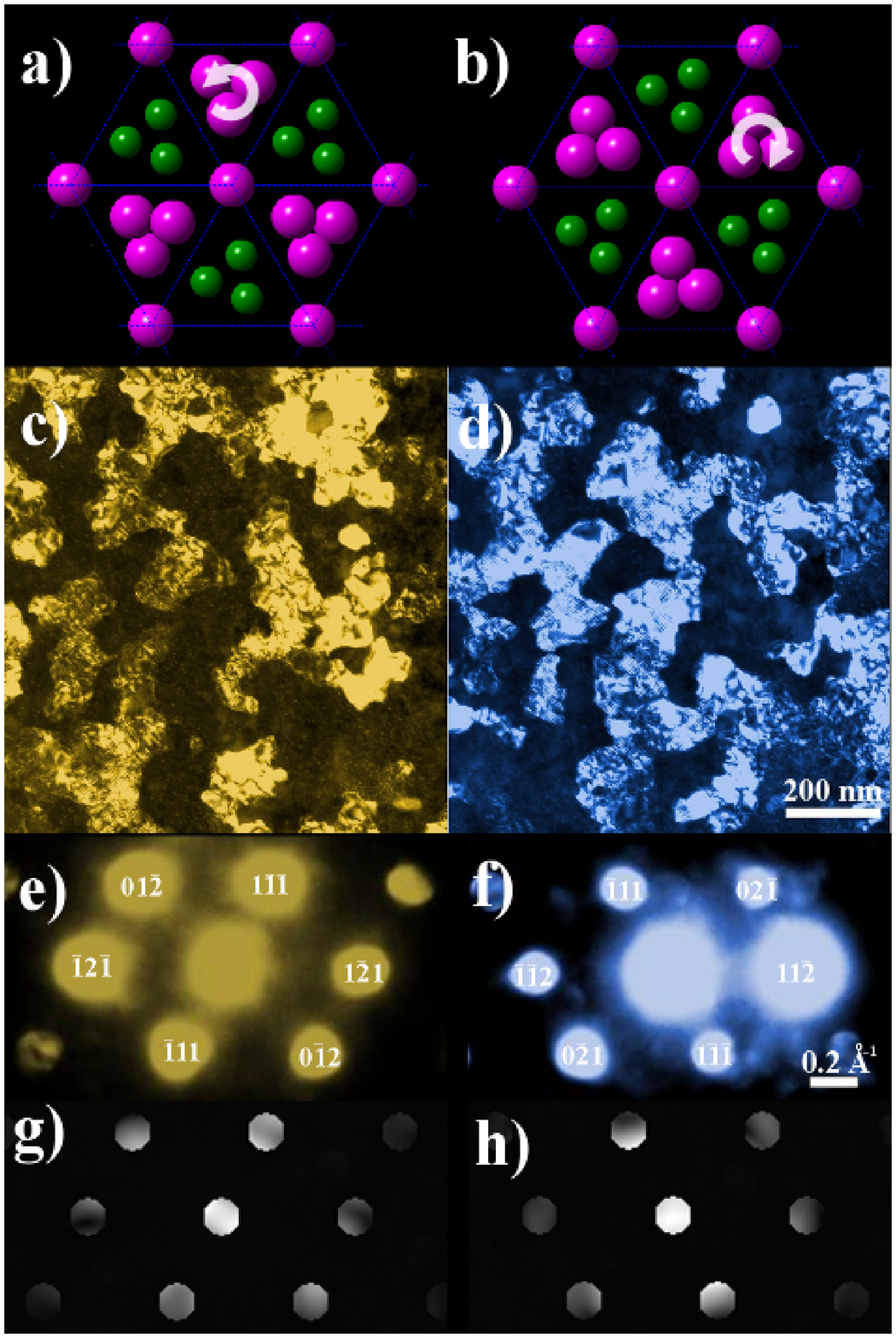}
  \caption{\textsf{\textbf{Chiral B20 crystal structure of \FeCoSi .} \textbf{a,} and \textbf{b,} Two possible B20 structures of opposite chirality viewed along a $\langle 111 \rangle$ direction, with large atoms relating to transition metal elements and small atoms to Si. \textbf{c,} and \textbf{d,} Dark field TEM images of an $x = 0.3$, 50~nm film viewed along the $[321]$ ZA, formed using the $(1\bar{1}\bar{1})$ and $(\bar{1}11)$ diffraction spots from figures \textbf{e,} and \textbf{f,}. These are associated with domains of opposite chirality. \textbf{e,} and \textbf{f,} Microdiffraction patterns from individual grains with opposite chiralities, which can be combined to produce the overall diffraction pattern shown in figure S1. \textbf{g,} and \textbf{h,} convergent beam microdiffraction simulations of \textbf{e} and \textbf{f}.}}\label{fig:TMS_TEM}
\end{figure}

The samples studied here were 50~nm thick films of Fe$_{0.7}$Co$_{0.3}$Si, grown epitaxially on Si (111) by molecular beam epitaxy (MBE) using the method described in references \onlinecite{Porter2012} and \onlinecite{Sinha2013}. These films appear to be single crystal on the basis of X-ray and low energy electron diffraction techniques\cite{Sinha2013}, but in fact show twinning structures that are subtle to detect. As the process of growing by MBE does not break the degeneracy between the two possible chiral structures of the B20 unit cell, one may expect grains of both left- and right-handed chirality, structures that are shown in figures \ref{fig:TMS_TEM}a and \ref{fig:TMS_TEM}b. It is possible to distinguish these two chiral grains using a dark field transmission electron microscopy (TEM) method\cite{Karhu2010}.  When viewing a plan-view specimen along the $[321]$ ZA, a non-coincident superposition of the two chiral structures is observed in the electron diffraction pattern (see figure S1 in the supplementary information).  Dark field images displayed in figures \ref{fig:TMS_TEM}c and \ref{fig:TMS_TEM}d were formed with the objective aperture centered over the $(1\bar{1}\bar{1})$ and $(\bar{1}11)$ reflections of figures \ref{fig:TMS_TEM}e and \ref{fig:TMS_TEM}f respectively. These reveal a structure comprised of opposite chirality grains, which are seen to be roughly equiaxed and typically 100-200 nm in lateral size. Microdiffraction patterns from areas of opposite chirality are shown in figures \ref{fig:TMS_TEM}e and \ref{fig:TMS_TEM}f, with convergent beam microdiffraction simulations of left-handed and right-handed  structures shown in \ref{fig:TMS_TEM}g and \ref{fig:TMS_TEM}h respectively. Left- and right-handed structures in bulk host magnetic helices of opposite handedness\cite{Grigoriev2009}. Hence, our epilayer will host chiral textures with opposite in-plane windings in equal abundance, with complex magnetisation textures to be found at the grain boundaries\cite{Yu2011}.

\begin{figure}[t]
  \includegraphics[width=8cm]{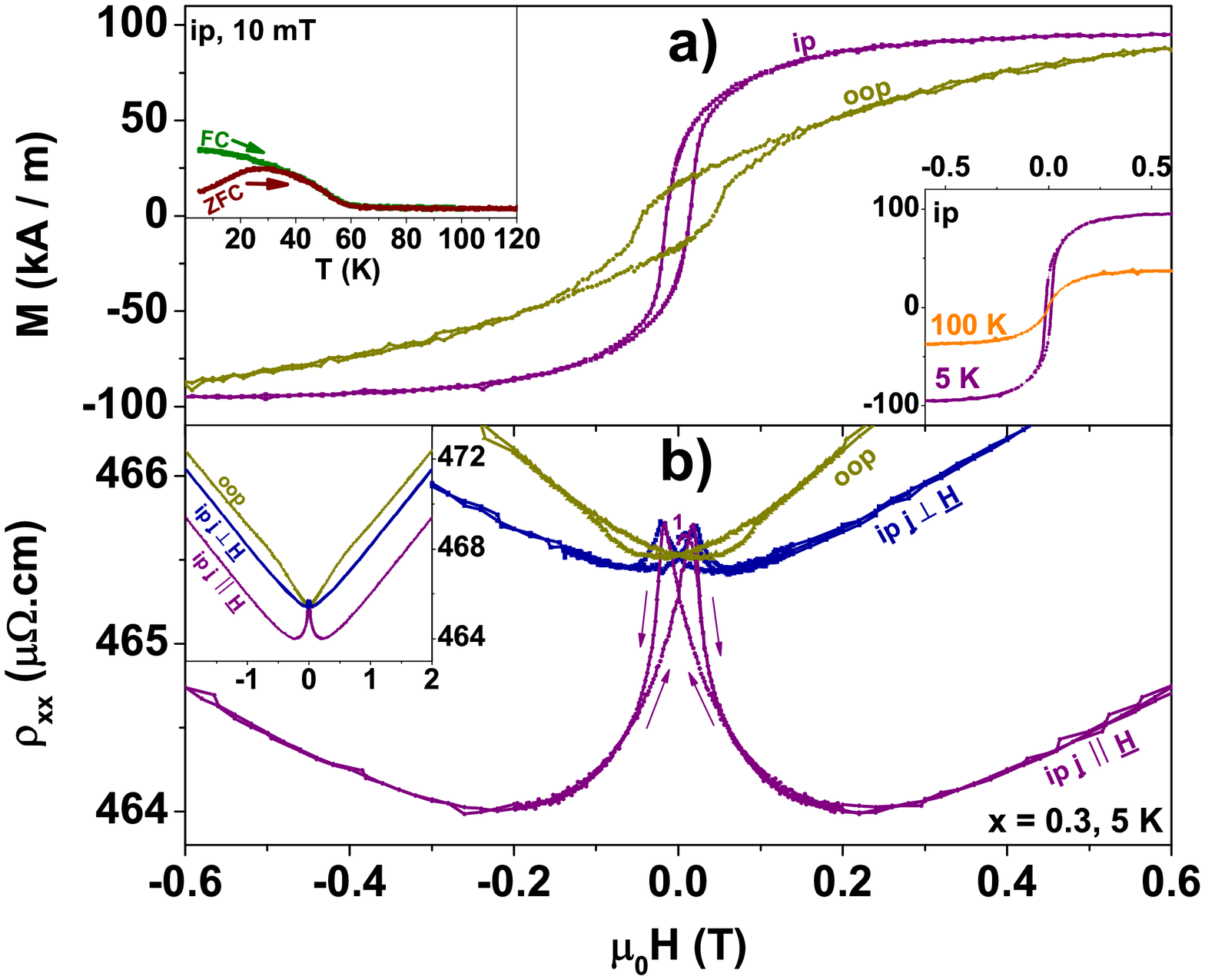}
  \caption{\textsf{\textbf{Anisotropy in a thin film of Fe$_{0.7}$Co$_{0.3}$Si.} \textbf{a,} Magnetometry at 5~K reveals an in-plane (ip) anisotropy with an out-of-plane (oop) hard axis, with hysteresis in both field orientations. Additional in-plane data at 100~K (inset bottom right) shows reduced hysteresis. ZFC and FC magnetisation, measured at 10~mT, is shown inset top left. \textbf{b,} The change in longitudinal resistivity at 5~K is anisotropic at low fields, shown here for out-of-plane (oop) and in-plane with field $H$ parallel (ip~$j \parallel H$) or perpendicular (ip~$j \perp H$) to the current density $j$. An isotropic linear contribution is observed for higher fields (shown inset).} \label{fig:MBE2600_anisotropy}}
\end{figure}

In bulk films the helical orientation is defined by weak anisotropic exchange along a cubic [100] axis\cite{Grigoriev2007}. In our films the $\sim 6 \%$ biaxial strain due to the lattice mismatch with the substrate is relaxed over 50 nm to $\sim 3\%$, distorting the unit cell to a rhombohedral form\cite{Sinha2013}. This epitaxial strain raises the magnetic ordering temperature\cite{Karhu2010,Porter2012,Li2013,Sinha2013}, in this case to $T^\mathrm{ord}_\mathrm{epi} \sim 60$~K relative the the bulk $T^\mathrm{ord}_\mathrm{bulk} \sim 45$~K\cite{Grigoriev2007}, as shown by the field cooled (FC) and zero field cooled (ZFC) magnetisation measurements in the inset of figure \ref{fig:MBE2600_anisotropy}a. The lattice distortion introduces a magnetic uniaxial anisotropy, revealed in the SQUID magnetisation loops shown in figure \ref{fig:MBE2600_anisotropy} for in-plane and out-of-plane magnetic fields $H$. By applying the model for helices in a film with a propagation vector out of the plane given by Karhu \emph{et al.}\cite{Karhu2012}, an easy-plane uniaxial anisotropy was found, with a magnetocrystalline anisotropy constant of $30 \pm 2$~kJm$^{-3}$.

This anisotropic response is also found in the magnetoresistance (MR). Figure \ref{fig:MBE2600_anisotropy}b shows the $H$ dependence of the longitudinal resistivity $\rho_{xx}$ for the three principal orientations: with $H$ out of plane, and with $H$ in-plane and the current parallel (~$j \parallel H$) or perpendicular (~$j \perp H$) to the field. At low fields a negative MR is observed, which is greatest for the ~$j \parallel H$ case. At high fields ($\mu_0 H \gtrsim 0.3$~T, shown inset) we observe the isotropic linear positive MR that is characteristic of B20 \FeCoSi \cite{Manyala2000,Onose2005,Porter2012,Sinha2013}.

The hysteresis in these epilayers is to be contrasted with bulk crystals, where a helical ground state can be unwound reversibly as a function of field\cite{Butenko2010} although hysteresis has been observed in the neutron scattering profile\cite{Grigoriev2007}. In these epilayers this is not the case, loss of the ZFC state is irreversible, unless the sample is warmed above $T^\mathrm{ord}_\mathrm{epi}$.

\section*{Chiral magnetism in \FeCoSi\ epilayers}

\begin{figure}[t]
  \includegraphics[width=8.5cm]{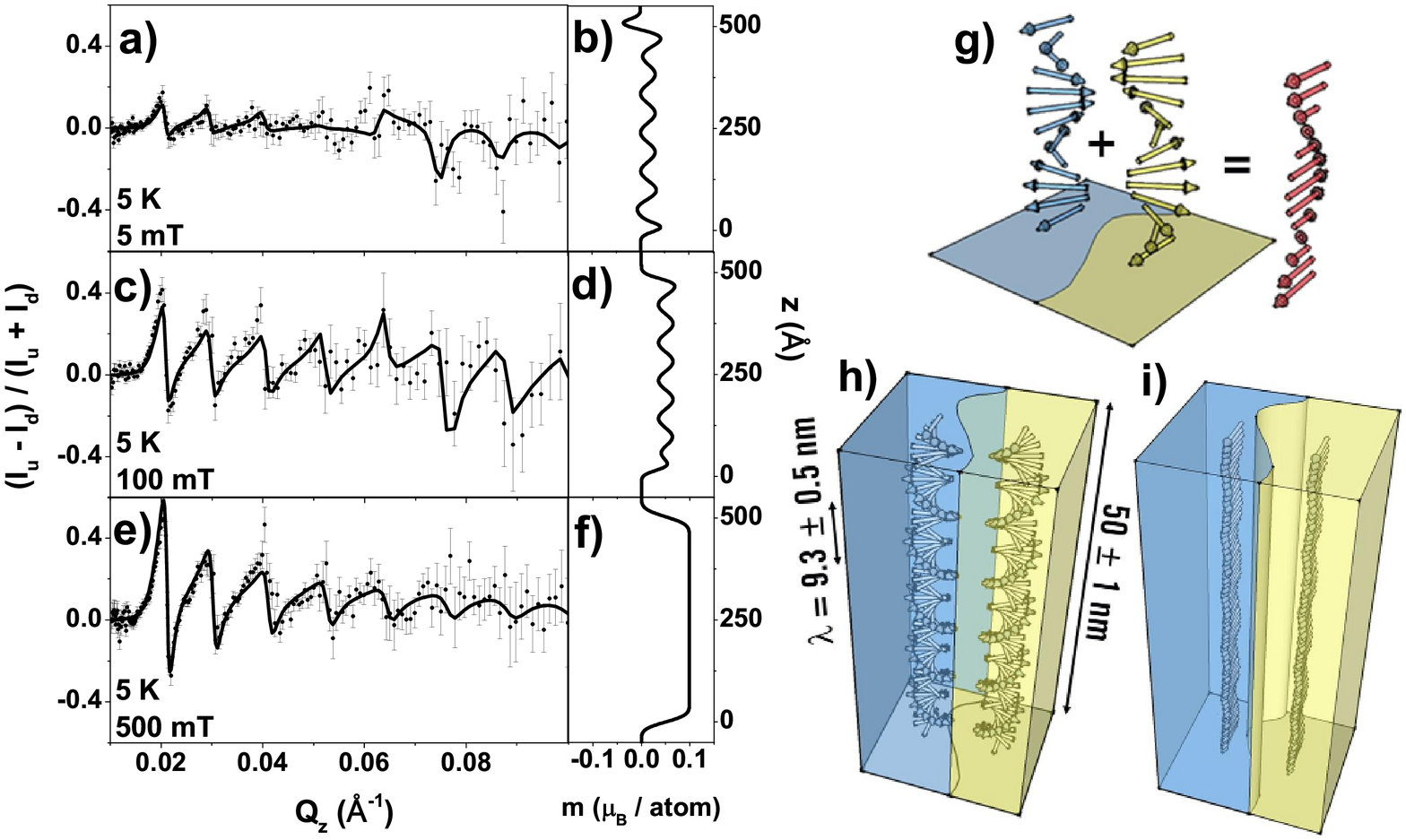}
  \caption{\textsf{\textbf{Chiral magnetism in an \FeCoSi\ epilayer.} PNR spin asymmetry (symbols) at 5~K after cooling in a 5~mT in-plane magnetic field is shown at \textbf{a,} 5~mT, \textbf{c,} 100~mT,  and \textbf{e,} a saturating field of 500~mT. Model SLDs are shown in \textbf{b}, \textbf{d}, and \textbf{f}, leading to the fits (lines) shown through the PNR data. \textbf{g,} The periodic spin density wave used to fit the data is a superposition of two distorted helices with opposite chirality aligned by the cooling field. \textbf{h,} The spin helices that lead to the SLD profile shown in \textbf{b}. \textbf{i,} The saturated state leading to the SLD profile shown in \textbf{f}.} \label{fig:SpinAsym}}
\end{figure}

The period of the helical wave in the \FeCoSi\ series has been studied in bulk using small angle polarised neutron scattering\cite{Grigoriev2009}. For $x = 0.3$, as used in this study, $\lambda_\mathrm{bulk} \approx 42$~nm. Based upon the easy-plane anisotropy, one may expect that there exists a helimagnetic ground state in these epilayers that can only be reached by cooling in fields close to zero. In order to determine the periodicity of the helices we performed polarised neutron reflectometry (PNR) on a larger epilayer (see supplementary figures S2 and S3 for its characterisation). The sample's easy plane implies in-plane spins and hence an out-of-plane propagation vector for the helices. Following the method of Karhu \textit{et al.}\cite{Karhu2011,Karhu2012}, the epilayer was cooled in a 5~mT in-plane field (smaller than the coercive field) which will align the helices with opposite chirality that will be found in the racemic grain structure determined by TEM (figure \ref{fig:TMS_TEM}). This field cooling then produces a spin density wave (SDW) as the spatial average of the helices of opposite chirality, shown schematically in figure \ref{fig:SpinAsym}g, which may be detected by PNR.

The spin asymmetry in the neutron reflectivities is shown for $H=5$, 100, and 500~mT (saturation) in figures \ref{fig:SpinAsym}a, \ref{fig:SpinAsym}c and \ref{fig:SpinAsym}e, respectively (see figure S4 in the supplementary information for the corresponding reflectivity spectra). The solid lines are fits generated from the scattering length density (SLD) profiles shown in figures \ref{fig:SpinAsym}b, \ref{fig:SpinAsym}d and \ref{fig:SpinAsym}f. At saturation the spin asymmetry is fitted well by a model with a uniform magnetic depth profile (figure \ref{fig:SpinAsym}i) with 0.10 $\mu_{B} / \mathrm{atom}$, in good agreement with the 0.12 $\mu_{B} / \mathrm{atom}$ calculated from the saturation moment measured by SQUID (figure \ref{fig:MBE2600_anisotropy}). At 5 and 100~mT a feature is observed at a wavevector transfer $Q_{z} \sim 0.075$ \AA$^{-1}$ that is not present at saturation. The position and sign of this (and the neighbouring) feature requires a periodicity of the magnetic profile of $\lambda_\mathrm{epi} = 9.3 \pm 0.5$ nm to achieve a good fit.

Combining this result with those from the magnetometry reveals differences in $SD_\mathrm{epi} = 1.6 \pm 0.2$~meV\AA\ compared to $SD_\mathrm{bulk} = 1.11$ meV\AA\ where $S$ is the spin per unit cell and $D$ the DMI constant. The spin wave stiffness, $A_\mathrm{epi} = 24 \pm 4$~meV\AA$^{2}$, was also changed from the bulk value\cite{Grigoriev2007}, $A_\mathrm{bulk} \sim 63$~meV\AA$^{2}$. The ratio $A / SD$ determines the spatial scale of magnetic modulations in the material.

The SLD profile at 5~mT arises from a model where the helices are distorted by the finite field\cite{Karhu2012}, depicted in figure \ref{fig:SpinAsym}h. This distortion will be greater at 100~mT. The SLD is offset from zero field implying a non-trivial chiral structure with periodicity $\lambda_\mathrm{epi}$, which might be a superposition of FM and periodic states, helicoids, a skewed conical phase, or in-plane skyrmions\cite{Karhu2012}. It is currently unclear what magnetic textures are expected at structural chiral grain boundaries and these are likely to be important in determining the true three-dimensional magnetic structure.


\section*{Topological contributions to magnetotransport}

\begin{figure*}[t]
  \includegraphics[width=12cm]{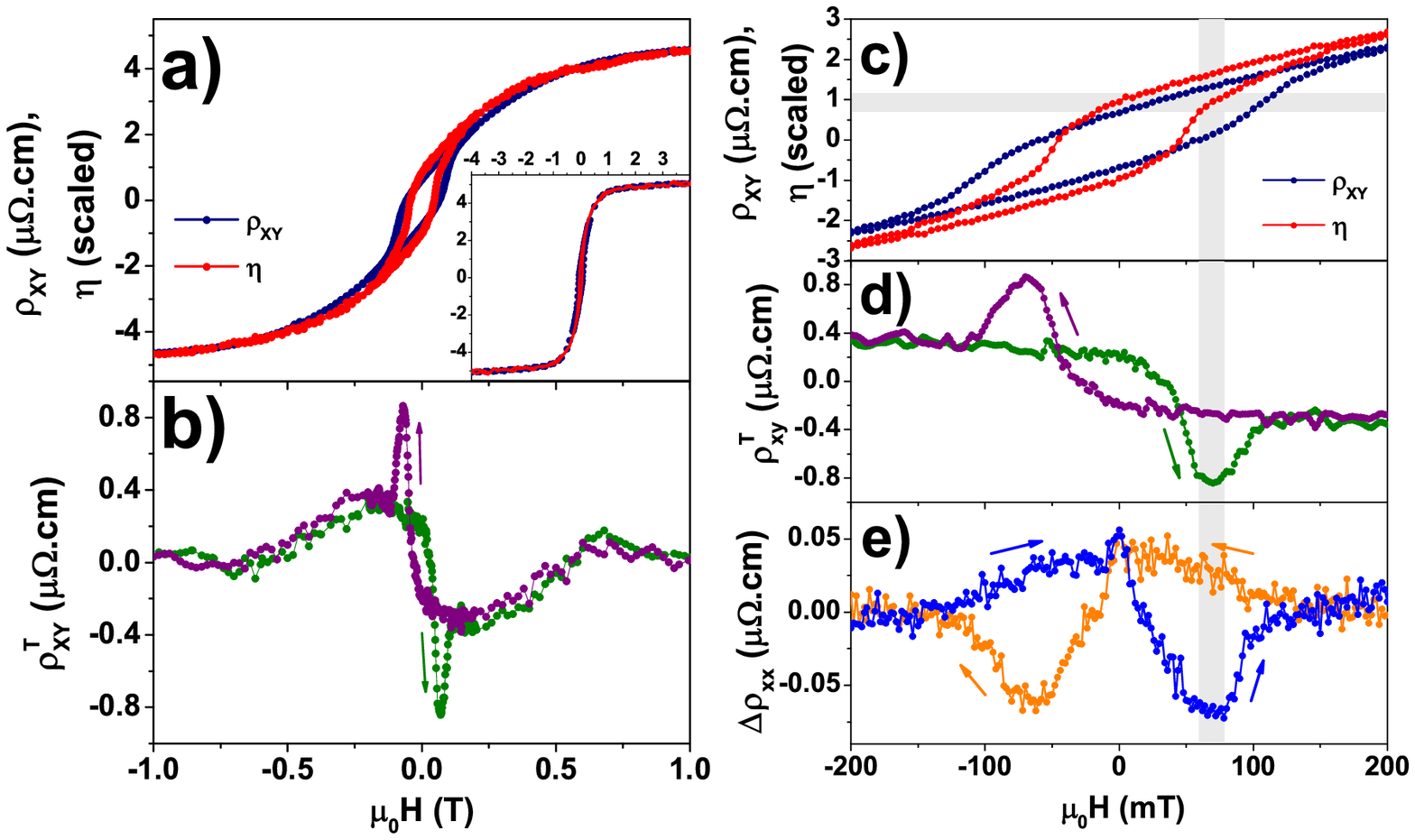}
  \caption{\textsf{\textbf{Transport and magnetic properties at 5 K for a 50 nm Fe$_{0.7}$Co$_{0.3}$Si film.} \textbf{a,} The Hall resistivity and scaled magnetisation are shown as a function of applied magnetic field $H$. Both measurements, extended to higher fields, are shown inset. \textbf{b,} The difference between the curves in \textbf{a} is interpreted as the THE contribution $\rho_{xy}^\mathrm{t}$. \textbf{c,d} Low field plots of \textbf{a} and \textbf{b} showing $\rho_{xy}^\mathrm{t}$ peaking at $\pm 65$~mT. This field corresponds to a non-zero magnetisation prior to saturation as indicated by the grey bars for a negative to positive field sweep. \textbf{e,} The residual longitudinal resistivity shows minima that coincide with the coercive field of neither $\rho_{xy}$ nor $\eta$, but with the extrema in $\rho_{xy}^\mathrm{t}$.} \label{fig:TMS_THE005K}}
\end{figure*}

Magnetic skyrmion states can be imaged directly\cite{Yu2010,Yu2011}, observed indirectly using scattering experiments\cite{Muhlbauer2009,Jonietz2010}, or have often been inferred from the influence of the chiral spin structure on conduction spins through the topological Hall effect\cite{Neubauer2009,Lee2009,Huang2012,Kanazawa2011}. As an electron traverses a magnetisation texture with its spin should adiabatically follow the local spin orientation. In doing so it accumulates a geometric (Berry) phase if the texture possesses a non-zero skyrmion winding number density, defined as $n_\mathrm{s} = \frac{1}{4 \pi} \boldsymbol{\hat\mathrm{M}} \cdot \frac{\partial \boldsymbol{\hat\mathrm{M}}}{\partial x} \times  \frac{\partial \boldsymbol{\hat\mathrm{M}}}{\partial y}$, where $\boldsymbol{\hat\mathrm{M}}$ is a unit vector along the local magnetisation direction: for a chiral skyrmion lattice integrating this over a unit cell yields exactly $-1$ per skyrmion. This phase can be considered to be equivalent to an Aharonov-Bohm phase in a fictitious magnetic field equal to minus one flux quantum, $-\Phi_{0}$, per skyrmion winding number\cite{Schulz2012}, leading to the so-called topological Hall effect (THE).

The Hall resistivity $\rho_{xy}$ in our films arises from three main contributions: the ordinary, anomalous, and topological Hall effects. Thus, we write the total Hall resistivity as:
\begin{equation}
  \rho_{xy}(H) = \rho_{xy}^\mathrm{o}(H) + \rho_{xy}^\mathrm{a}(H) + \rho_{xy}^\mathrm{t}(H).\label{eqn:AllHallSupp}
\end{equation}
From the ordinary Hall contribution $\rho_{xy}^{o}(H) = R_{0}H$, where $R_{0}$ is the ordinary Hall coefficient, we can determine the film carrier density, $n = -1/R_0 e$. The anomalous Hall resistivity $\rho_{xy}^\mathrm{a}(H) = R_\mathrm{s} M(H)$ originates from spin-orbit coupling and is proportional to the magnetisation, $M(H)$\cite{Nagaosa2010}. $R_\mathrm{s}$ is a function of the longitudinal resistivity $\rho_{xx}$ and thus varies with temperature and field. The final term is the topological Hall contribution $\rho_{xy}^\mathrm{t}$.

The measured total Hall resistivity $\rho_{xy}(H)$ of our film at 5 K is shown in figures \ref{fig:TMS_THE005K}a and \ref{fig:TMS_THE005K}c, and exhibits hysteresis. In the same figure we plot a scaled form of the  magnetisation $\eta(H) = S_\mathrm{H}\rho_{xx}^{2}(H) M(H) + R_{0}H$. The meaning of the scaling constant $S_\mathrm{H}(H)$ and the scaling procedure are described in the supplementary information. Inspection of equation \ref{eqn:AllHallSupp} shows that $\rho_{xy}^\mathrm{t}(H) = \rho_{xy}(H) - \eta(H)$, which is plotted in figures \ref{fig:TMS_THE005K}b and \ref{fig:TMS_THE005K}d. This displays sharp extrema at $\mu_0 H = \pm 65$ mT superposed on a weakly hysteretic broad background extending up to about 0.7~T. The sharp features occur at a field that corresponds to a non-zero magnetisation where a kink appears in $M(H)$.

The THE arises from the presence of magnetic textures with non-zero skyrmion winding number, and skyrmions have been observed in \FeCoSi\ at equivalent fields\cite{Yu2010}, so it seems natural to assume that the observed THE is due to the formation of skyrmions in our epilayers. Whilst there are, in addition to this work, a few reported instances of the observation of a topological Hall effect in B20-ordered epilayers of various materials under these conditions\cite{Huang2012,Li2013,Wilsonthesis2013}, recent analytical modelling predicts that an easy-plane anisotropy should suppress the formation of skyrmions in an out-of-plane field\cite{WilsonUnpub2013}. It may be that the approximations in the analytical model do not fully capture the relevant physics and these results can be accounted for by the presence of true chiral skyrmion lattice textures. On the other hand, Wilson \textit{et al.} proposed a laterally modulated cone phase texture\cite{Wilsonthesis2013} that also has non-zero skyrmion winding number. Various materials have exhibited the THE, the majority of which are B20 crystals, but with some exceptions such as orthorhombic (B31) MnP which has a fan like spin structure\cite{Shiomi2012} implying that magnetic textures other than the skyrmions may generate a THE. It is therefore convenient to discuss our results in terms of an areal skyrmion winding number density $n_\mathrm{s}$. In the case of a true skyrmion lattice, $n_\mathrm{s}$ is the skyrmion areal number density.

The peak magnitude of the THE we observe, $\rho_{xy}^\mathrm{t} \approx 820$~n$\Omega$cm, is very large. A recent study of MnSi under pressure reported a giant THE of $\sim 50$~n$\Omega$cm\cite{Ritz2013}, roughly an order of magnitude greater than under ambient conditions\cite{Neubauer2009}. Of comparable size is the THE in the cubic perovskite SrFeO$_3$ \cite{Ishiwata2011}. The largest effect observed prior to our observations in Fe$_{0.7}$Co$_{0.3}$Si was in MnGe (160 n$\Omega$cm)\cite{Kanazawa2011}, which has very small (4~nm) skyrmions, implying an emergent magnetic field of hundreds of teslas.

Using a simple model\cite{Huang2012}, we can account for the very large value of the THE in our Fe$_{0.7}$Co$_{0.3}$Si epilayer. A large THE requires a large spin polarization of the current, $P$, a high areal density of skyrmion winding number, $n_\mathrm{s}$, and a low carrier concentration. Since the emergent field is simply $-n_\mathrm{s} \Phi_0$, the THE can be written as
\begin{equation}
  \rho_{xy}^\mathrm{t} = -n_\mathrm{s} \Phi_{0} P  R_{0}.\label{eqn:THE}
\end{equation}
Fe$_{0.7}$Co$_{0.3}$Si is a doped semiconductor with the parent (FeSi) a narrow gap insulator\cite{Manyala2000}. By virtue of this doping, the magnitude of $R_{0}$ for our sample is $64 \pm 4$~n$\Omega$cmT$^{-1}$ (corresponding to approximately one electron-like carrier per Co dopant), much larger than metallic MnSi\cite{Schulz2012} (16~n$\Omega$cmT$^{-1}$). Electron doping with cobalt introduces carriers which for $0.05 < x < 0.25$ are fully spin polarised\cite{Manyala2000,Guevara2004}. Our $x = 0.3$ films have been measured to have a high spin polarisation of $P = 0.77$\cite{Sinha2013}. The high spin polarisation and ordinary Hall coefficient contribute towards the large THE we observe.

The last remaining factor to address is the emergent field: the values given above imply a value of $B_\mathrm{eff} = -n_\mathrm{s} \Phi_0 \approx -17$~T at the THE peak. Using the PNR helical wavevector and considering a hexagonally close-packed skyrmion crystal, a skyrmion size of $2\lambda_\mathrm{epi}/ \sqrt{3} = 10.7 \pm 0.6$~nm is obtained. Substituting the above parameters into equation \ref{eqn:THE} requires $n_\mathrm{s} / n_\mathrm{Hx} = 0.2$ to match the observed magnitude of $\rho_{xy}^\mathrm{t}$, where $n_\mathrm{Hx}$ is the density for a hexagonally close-packed skyrmion crystal of the expected size. Assuming the presence of skyrmions, there are three possible explanations for this discrepancy: a much lower skyrmion packing density than $n_\mathrm{Hx}$ (skyrmions in a MnSi epilayers were reported to take up a glassy arrangement\cite{Li2013}); anisotropy in the helix pitch due to the rhombohedral crystal structure, yielding larger in-plane skyrmions than the measured out-of-plane helix period implies; or regions that lack a topological winding number (skyrmions are not supported close to the chiral grain boundaries\cite{Yu2011}). All three effects are likely to co-exist in our samples.

\begin{figure}[t]
  \includegraphics[width=8cm]{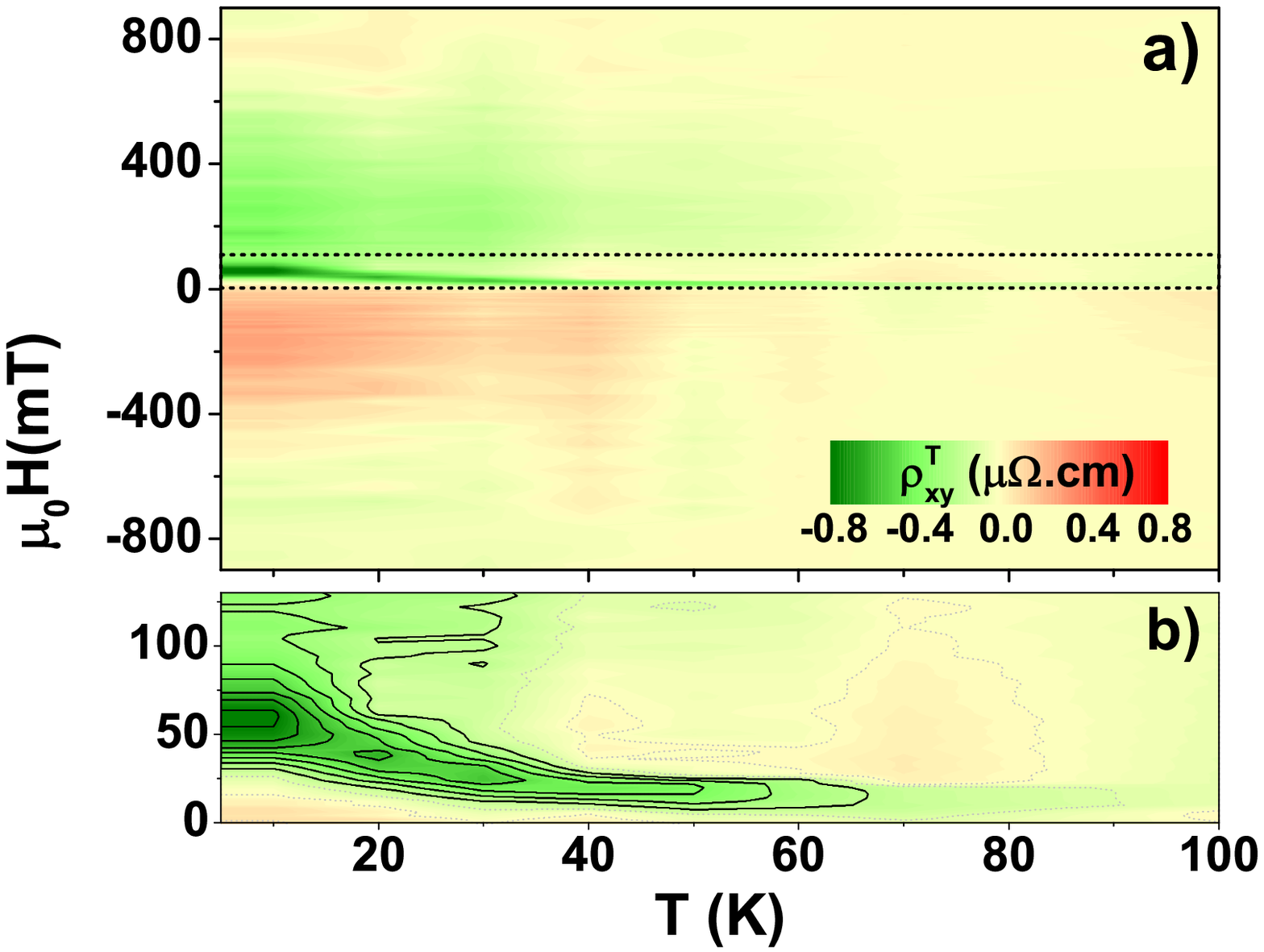}
  \caption{\textsf{\textbf{Temperature-field phase diagram.} The magnitude of the THE for a 50 nm thick Fe$_{0.7}$Co$_{0.3}$Si epilayer. Contour maps were interpolated for curves taken at 2, 5, 10, 20, 30, 40, 45, 50, 60, 70, and 100~K. Only sweeps from negative to positive fields are shown. A magnified view of the dashed box highlighted in \textbf{a} is shown in \textbf{b}.}\label{fig:TMS_THEmaps}}
\end{figure}

The temperature dependence of the THE is illustrated in figure \ref{fig:TMS_THEmaps}, where we show a phase diagram that has been constructed from a series of isothermal $\rho_{xy}(H)$ and $M(H)$ sweeps (see supplementary figure S6). The THE persists from $T^\mathrm{ord}_\mathrm{epi} \approx 60$~ K down to the lowest measurement temperature. The sharp maximum in THE is negative for all measured temperatures. This observation is in agreement with the negative effective field in conjunction with a positive $R_{0}$. Equation \ref{eqn:THE} then implies Fe$_{0.7}$Co$_{0.3}$Si has a positive spin polarisation and one would expect this behaviour where the majority spin carriers are at the Fermi energy where there is relative absence of minority spins\cite{Guevara2004}.

The two contributions to the THE are illustrated by figures \ref{fig:TMS_THEmaps}a and \ref{fig:TMS_THEmaps}b. The sharp, hysteretic maximum in the THE is shown in the latter, and the phase diagram is reminiscent of that for a free skyrmion crystal, as described for bulk material, occurring at magnetic fields (tens of mT) similar to those reported for unstrained samples\cite{Yu2010}. Figure \ref{fig:TMS_THEmaps}a shows the broader contribution similar to that observed in thin films\cite{Huang2012,Li2013} extending to $\sim$ 700 mT. At these higher fields the THE and effective flux is smaller and may be attributed to chiral defects with a topological winding number, such as pinned skyrmions that might be expected at the chiral grain boundaries, persisting to higher fields. Nevertheless, the topologically non-trivial phase extends over all $T < T^\mathrm{ord}_\mathrm{epi}$ and a broad field range.

Our data also hint at the detection of a topological contribution to $\rho_{xx}$. After a linear background and a small contribution from 2D weak localisation\cite{Porter2012} were removed from the out-of-plane MR, a residual contribution, $\Delta\rho_{xx} (H)$, remains, as shown in figure \ref{fig:TMS_THE005K}e. Hysteretic minima are seen. Crucially, these minima do not align with the coercive field determined from the magnetisation shown in figure \ref{fig:TMS_THE005K}d, and so cannot be attributed to anisotropic magnetoresistance. Instead they align with the sharp, hysteretic extremum in the THE. This contribution to the magnetoresistance, $\Delta\rho_{xx} (H) \approx 100$~n$\Omega$cm, is about an order of magnitude smaller than $\rho_{xy}^\mathrm{t}$. Contributions to the conductivity from topological Berry phase effects are so far formally predicted only for $\rho_{xy}$, but Monte Carlo simulations by Yi \emph{et al.} have suggested that traces of these effects may also be observed in the longitudinal resistivity\cite{Yi2009}. In that report, a rise in $\rho_{xx}$ was simulated that may be attributed to an enlargement of the magnetic unit cell, but no firm conclusion on its origin were drawn. Weak kinks in $\rho_{xx}$ were observed in an early study of MnSi under pressure\cite{Lee2009}. In contrast, we observe a drop in longitudinal resistivity in the field range where skyrmions are present. In order to interpret the origin of this effect, a complete knowledge of the magnetic topology may be required. This contribution is lost for $T \gtrsim 20$ K (see supplementary Fig. S7), suggesting that phase coherence may play a role in the underlying mechanism.

\section*{In Conclusion}

A large THE has been observed in Fe$_{0.7}$Co$_{0.3}$Si epilayers by virtue of the high spin-polarisation, low carrier concentration, and compact skyrmion size. Fingerprints of possible topological contributions are also observed in the longitudinal resistivity. Many of these parameters, such as doping level or epitaxial strain, are open to engineering in order to give control of the functional properties arising from these novel topological spin textures. Our epilayers are grown on commercial Si wafers and are amenable to planar processing required for device manufacture. The presence of such a large THE has important implications for skyrmion-based spintronics\cite{Kiselev2011,Schulz2012,Sampaio2013,IwasakiNANO2013}, such as the detection of skyrmions in current-controlled memory cells\cite{Lin2013}.

\section*{Methods}

\footnotesize{
Magnetometry, transport and all characterisation (transmission electron microscopy, X-ray diffraction and low energy electron diffraction) except for PNR were all performed on pieces cut from a sample with composition $x = 0.3$ from the same starting wafer. Hall bars that were defined by standard photolithography with 5~$\mu$m spacing were used for measurements of both the Hall resistivity, $\rho_{xy}$, and the longitudinal resistivity, $\rho_{xx}$, using standard 4-wire dc techniques. For both measurements a current density of $j = 4 \times 10^{8}$~Am$^{-2}$ was used, but the results were insensitive to $j$ down to values of $j = 2 \times 10^{4}$~Am$^{-2}$. SQUID-VSM was used to measure the magnetisation, $M$, as a function of applied field, $H$, over a range of temperatures from 2-300~K. Van der Pauw measurements of $\rho_xy$ on a sheet film were made to rule out the possibility that the observed THE is an artifact caused by a change in coercivity in the Hall bar during patterning. Plan-view TEM sections of \FeCoSi\ films were prepared by focused ion beam (FIB) milling and imaged using a Philips CM200 FEGTEM. TEM convergent beam microdiffraction simulations were performed using the multislice method\cite{Kirkland1998}, a beam convergence of 1.8 mrad, an incoherent sum of images from 64 thermal vibration configurations, and a 50 nm thick film with 30 \% random substitution of Co for Fe in the silicide.

A separate 20~mm $\times$ 20~mm sample (identically prepared) was used for PNR study and had properties very close to the specimen used for other characterisation methods (see supplementary information). All PNR measurements were made after heating the sample above the ordering temperature to 100 K and cooling in a magnetic field of 5 mT to 5 K. 5, 100 and 500 mT measurements were performed sequentially, using the PolRef reflectometer at ISIS. The data were fitted using the \texttt{GenX} software\cite{Bjorck2007}.
}


\vskip 2cm
Correspondence and requests for materials should be addressed to C.H.M.

\section*{Acknowledgments}
We are grateful to G. Tatara, S. Huang, and T. L. Monchesky for useful discussions. This work was supported financially by EPSRC (grant number EP/J007110/1) and the European Union through the Marie Curie Initial Training Network ``Q-NET''. Experiments at the ISIS Pulsed Neutron and Muon Source were supported by a beamtime allocation from the Science and Technology Facilities Council.

\section*{Author Contributions}
N.A.P and P.S. grew and characterised the epilayers; N.A.P. patterned the Hall bars and made the magnetotransport measurements; M.B.W. made the TEM specimens and performed the TEM analysis with N.A.P. under the supervision of R.M.D.B.; M.D.R. performed TEM CBED simulations; N.A.P and P.S. performed the PNR experiments and fitted the data with the assistance of T.R.C., C.J.K., and S.L.; N.A.P. and A.N.D. performed the SQUID-VSM measurements. N.A.P. analysed the magnetotransport data; C.H.M. and N.A.P. proposed the study and wrote the paper; all authors discussed the data and commented on the manuscript.

\section*{Additional Information}
The authors declare no competing financial interests. Supplementary information
accompanies this paper on www.nature.com/naturematerials. Reprints and permissions information is available online at http://npg.nature.com/reprintsandpermissions. Correspondence and requests for materials should be addressed to N.A.P. or C.H.M.

\end{document}